\input{aipcheck}

\documentclass[,final         
]
  {aipproc}

 \layoutstyle{6x9}
 \usepackage{color}
 \usepackage{amsmath,amssymb,mathrsfs}
 \DeclareSymbolFont{symbols}{OMS}{cmsy}{m}{n}

 \def\a{{\alpha}}
 \def\b{{\beta}}
 
\def\q{{\textrm{q}}}

 \def\z{{\zeta}}

 \def\bx{{\mathbf x}}
 \def\bp{{\mathbf p}}
 \newcommand{\ben}{\begin{eqnarray}}
 \newcommand{\een}{\end{eqnarray}}
 \def\be{\begin{equation}}
 \def\ee#1{\label{#1}\end{equation}}

\begin{document}

\title{Relativistic Ohm and Fourier laws for binary mixtures of electrons with protons and photons}


\author{Gilberto M. Kremer}{
              address={Departamento de F\'{\i}sica, Universidade Federal do Paran\'a, Curitiba, Brazil}}



\begin{abstract}
Binary mixtures of
electrons with protons and of electrons with photons
subjected to external electromagnetic fields are analyzed by using the Anderson and
Witting model equation.
The relativistic laws of Ohm and Fourier
are determined as well as general expressions
for the electrical and thermal conductivities for
relativistic ionized gas mixtures.
Explicit expressions for the transport
coefficients are given for
the particular cases:
 a non-relativistic mixture of protons and
non-degenerate electrons;
 an ultra-relativistic mixture of photons and non-degenerate electrons;
 a  non-relativistic mixture of protons and completely degenerate
electrons;  an ultra-relativistic mixture of photons and completely
degenerate  electrons and  a mixture of non-relativistic protons and
ultra-relativistic completely degenerate  electrons.

\end{abstract}

\maketitle

\section {Introduction}

The analysis of non-relativistic and relativistic ionized gases
by using the Boltzmann equation is a very difficult subject,
since it refers to a system of coupled nonlinear  integro-differential equations
for the distribution functions.  Simpler model equations for the
collision term have been proposed in the literature
in order to overcome the difficulties of the Boltzmann integro-differential equation. The model equations  simplify   the  structure of the collision term but  maintain
its basic properties.  For the non-relativistic
Boltzmann equation the most widely known model is the  BGK model
which was formulated independently by Bhatnagar,
Gross and Krook~\cite{BGK} and Welander~\cite{WE}. The first extention of the
non-relativistic BGK model to the relativistic case was proposed by Marle~\cite{MAR}. Although the non-relativistic limiting case of the Marle's model recovers the non-relativistic BGK model,  in
the case of particles with zero rest mass the
relaxation time of the distribution function  tends to infinity. This shortcoming was found by
 Anderson and Witting~\cite{AW} who proposed a new model equation.

In this work we follow~\cite{KP} and analyze binary mixtures of
electrons and protons
and of electrons and photons
subjected to external electromagnetic fields within the framework of Anderson and
Witting model equation.
These two systems are important in astrophysics
since they could describe
magnetic white dwarfs or
cosmological fluids in the plasma period and in the
radiation dominated period. By using the Chapman-Enskog methodology we determine Ohm and Fourier  laws in the presence of electromagnetic fields
and general expressions for the electrical and thermal conductivities for
relativistic non-degenerated and degenerate binary mixtures of electrons with
protons and electrons with photons. Furthermore, explicit expressions for these
coefficients are given for the particular mixtures:
(a) a non-relativistic mixture of protons and
non-degenerate electrons;
(b) an ultra-relativistic mixture of photons and non-degenerate electrons;
(c) a  non-relativistic mixture of protons and completely degenerate
electrons; (d) an ultra-relativistic mixture of photons and completely
degenerate  electrons and (e) a mixture of non-relativistic protons and
ultra-relativistic completely degenerate  electrons.

\section{Relativistic Uehling-Uhlenbeck equation}

Let us first consider a single relativistic quantum ideal gas in a Minkowski space
characterized by metric tensor $\eta^{\alpha\beta}$ with signature
diag(1,-1,-1,-1).
In the phase space spanned by the space-time coordinates $(x^\alpha)=(ct,x^\alpha)$ and momentum four-vector $(p^\alpha)=(p^0, {\bf p})$
the state of the relativistic quantum gas is characterized by the one-particle distribution function $f(x^\alpha, p^\alpha)\equiv f(\bx, \bp,t)$,
since the length of the momentum four-vector is given by $mc$ so that $p^0=\sqrt{\vert\bp\vert^2+m^2 c^2}$. The number of particles at time $t$ in the volume
element $d^3x$ about $\bf x$ and with momenta in the range $d^3p$ about ${\bf p}$ is given by $f({\bf x}, {\bf p},t)d^3xd^3p$.

The space-time evolution of the  one-particle distribution function
$f({\bf x},{\bf p},t)\equiv f$  in the phase space is given by the Boltzmann equation (see e.g.~\cite{GR8,CK})
\be
p^\alpha {\partial f\over \partial x^\alpha}
+m{\partial f K^\alpha \over \partial p^\alpha}
=\mathcal{Q},
\ee{8.127}
where $m$ denotes the rest mass of the particle and $K^\a$ is the Minkowski force which acts on the particles of the gas.
Furthermore, $\mathcal Q$ is a term which takes into account the collisions of the particles. For a relativistic gas which obeys the classical statistical mechanics it is given by
\ben
\mathcal{Q}=\int
\left(f_*'f'-f_*f\right) \,F\,\sigma\,d\Omega
{d^3p_*\over p_{*0}}.
\een
In the above equation we have introduced the abbreviations
$f_*'\equiv f({\bf x,p}_*',t),$ $f'\equiv f({\bf x,p'},t),$ $f_* \equiv f({\bf x,p}_*,t),$ $f \equiv f({\bf x,p},t),$ where $\bp$ and $\bp_*$ denote
 the momenta of two particles before a binary collision    and $\bp'$ and $\bp'_*$ are the corresponding momenta after collision. The pre and post collisional momentum four-vectors are connected by the energy-momentum conservation law $p^\a+p_*^\a=p^{\prime\a}+p_*^{\prime\a}$.  Furthermore, $F =\sqrt{(p^\alpha_*p_\alpha)^2-m^4c^4 }$ is the invariant flux, which in the non-relativistic limiting case is proportional to the modulus of the relative velocity. The differential cross-section and the element of solid angle that characterize the binary collision are denoted by $\sigma$ and $d\Omega$, respectively.

The collision term $\mathcal Q$ for a gas whose particles obey quantum statistics may be motivated as follows.
First we note that  the volume element in the phase space $d^3xd^3p$ is a scalar invariant, but when quantum effects
are taken into account in a semi-classical description, we divide the
volume element by $h^3$, where $h=6.626\times$ $10^{-34}$
J s is the Planck constant. Hence we write ${d^3x d^3p/ h^3}$,
which is also a scalar invariant.
The term  ${d^3x d^3p/ h^3}$ may be interpreted as the number of available
states in the volume element $d^3x d^3p$. For particles with spin
 $\rm s$ there are more states, corresponding to the values that the spin
component on a given axis can take and we have to introduce the
degeneracy factor $g_{\rm s}$. Hence  the
number of available states is given by
\be
g_{\rm s}{d^3x d^3p\over  h^3}\qquad\hbox{where}\qquad
g_{\rm s}=\left\{
            \begin{array}{ll}
              2{\rm s}+1 & \hbox{for} \qquad m\neq0; \\
              2{\rm s} & \hbox{for}\qquad m=0.
            \end{array}
          \right.
\ee{2n.13}

In quantum mechanics a system of identical
particles may be  described by two kinds of particles:
 bosons and fermions. Bosons have integral spin, obey the Bose-Einstein
statistics and include mesons (pion, kaon),
photons, gluons  and nuclei of even mass number like helium-4. Fermions have
half-integral spin, obey the Fermi-Dirac statistics and include
leptons (electron, muon, tau), baryons (neutron,
proton) and nuclei of odd mass number like helium-3.
The main difference between bosons and
fermions in quantum statistical mechanics refers to the occupation number of a state. Any number of
boson particles may occupy the same state, while fermion particles obey the
Pauli exclusion principle and at most one particle may occupy each state.

In order to incorporate the  statistics of bosons and
fermions into the collision term, we begin to analyze fermions and note that due
the Pauli exclusion principle, the phase space is completely occupied
if the number of the particles in $d^3xd^3p$ is equal
to the number of available states
$fd^3xd^3p=g_{\rm s}{d^3xd^3p/ h^3}$, so that $f={g_{\rm s}/h^3}$. Hence, $(1-fh^3/g_{\rm s})$ gives the number of vacant
states in the phase space.
If the number of particles that enter
the volume element $d^3xd^3p$ in phase space, as a
consequence of a binary collision, is proportional to $f'f'_*$ this quantity must be multiplied by the
number of vacant states which is proportional to $(1-fh^3/g_{\rm
s})(1-f_*h^3/g_{\rm s})$. Hence the
following substitution in the collision term of the
Boltzmann equation must be consider:
\be
f'f'_*\longmapsto f'f'_*\left(1-{fh^3\over g_{\rm s}}\right)
\left(1-{f_*h^3\over g_{\rm s}}\right).
\ee{2.n15}
On the basis of the same reasoning we have to substitute
\be
ff_*\longmapsto ff_*\left(1-{f'h^3\over g_{\rm s}}\right)
\left(1-{f'_*h^3\over g_{\rm s}}\right),
\ee{2.n16}
for the particles that leave the volume
element $d^3xd^3p$ in phase space.

To include the apparent attraction between the boson particles  -- due to the statistics of
indistinguishable particles with no restrictions on the occupation of a
state -- the factor $(1-fh^3/g_{\rm s})$ must be
replaced by $(1+fh^3/g_{\rm s})$ . Hence we can write from the Boltzmann
equation (\ref{8.127}) and from the above conclusions
the relativistic Uehling-Uhlenbeck equation (for the non-relativistic Uehling-Uhlenbeck equation see~\cite{UU})
\ben\nonumber
&&p^{\alpha}{\partial f\over \partial x^\alpha}
+m{\partial fK^{\alpha}\over \partial p^\alpha}
=\int
\Biggl[f_*'f'\left(1+\varepsilon{fh^3\over g_{\rm s}}\right)
\left(1+\varepsilon{f_*h^3\over g_{\rm s}}\right)
\\\label{2.n17}
&&-f_*f\left(1+\varepsilon{f'h^3\over g_{\rm s}}\right)
\left(1+\varepsilon{f'_*h^3\over g_{\rm s}}
\right) \Biggr]\,F\,\sigma\,d\Omega
{d^3p_*\over p_{*0}},
\een
where $\varepsilon$ is defined through
\be
\varepsilon=\left\{
              \begin{array}{ll}
                +1 & \hbox{for Bose-Einstein statistics;} \\
                -1 & \hbox{for Fermi-Dirac statistics;} \\
                0 & \hbox{for Maxwell-Boltzmann  statistics.}
              \end{array}
            \right.
\ee{2.n18}

At equilibrium the number of particles that enter and leave the volume element  in the phase space must be equal to each other, so that the quantity within the brackets in (\ref{2.n17}) must vanish. Equivalently,  $\ln\left[{f^{(0)}/( 1+\varepsilon{f^{(0)}h^3/ g_{\rm s}}})\right]$ must be a summational invariant -- i.e., a function that obeys the relationship $\psi+\psi_*=\psi'+\psi'_*$ -- where $f^{(0)}$ denotes the equilibrium distribution function. For summational invariants there exists the following theorem (see e.g. \cite{CK}):  A continuous and differentiable
function of class $C^2$ $\psi(p^\alpha)$ is a summational invariant if and only if it is given by $\psi(p^\alpha)=A+B_{\alpha}p^{\alpha}$,
where $A$ is an arbitrary scalar and $B_\alpha$ an arbitrary
four-vector that do not depend on $p^\alpha$. Hence we have
\be
\ln\left({f^{(0)}\over 1+\varepsilon{f^{(0)}h^3
/ g_{\rm s}}}\right)=-(A+B^\alpha p_\alpha),\quad\hbox{or}\quad
f^{(0)}={g_{\rm s}/h^3\over e^{-a+B^\alpha p_\alpha}-\varepsilon},
\ee{2.n30}
where $a=-A-\ln(g_{\rm s}/h^3)$.

For the determination of $a$ and $B^\a$ we refer to \cite{CK}. Here we give only the results that $a=\mu/kT$ and $B^\a=U^\a/kT$, where $\mu$ is the chemical potential, $T$ the temperature, $k$ the Boltzmann constant, and $U^\a$ the four-velocity (with $U^\a U_\a=c^2$). Hence, the equilibrium distribution function reads
\ben\label{tn40}
f^{(0)}={g_{\rm s}\over h^3} e^{{\mu\over kT}-{U^\alpha p_\alpha
\over kT}},\qquad
f^{(0)}={g_{\rm s}/h^3\over e^{-{\mu\over kT}+{U^\alpha p_\alpha\over kT}}
\pm 1},
\een
when $\epsilon=0$ and $\epsilon=\mp 1$, respectively.
The relativistic Maxwell-Boltzmann distribution function (\ref{tn40})$_1$ was  obtained by
J\"uttner~\cite{J1} in 1911 and the relativistic Fermi-Dirac $(+)$ and Bose-Einstein $(-)$ distribution function (\ref{tn40})$_1$ was deduced by him~\cite{J2}
in 1928.

The extension of the Uehling-Uhlenbeck equation to a mixture of $r$ constituents is straightforward. We introduce an one-particle distribution function for each constituent  of the mixture $f_a\equiv f(\bx,\bp_a,t)$ $( a=1,\dots,r)$ which must satisfy the equation
\ben\nonumber
&&p_a^{\alpha}{\partial f_a\over \partial x^\alpha}
+\frac{q_a}{c}F^{\a\b}p_{a\b}{\partial f_a\over \partial p_a^\alpha}
=\sum_{b=1}^r\int
\Biggl[f_b'f_a'\left(1+\varepsilon_a{f_ah^3\over g^a_{\rm s}}\right)
\left(1+\varepsilon_b{f_bh^3\over g^b_{\rm s}}\right)
\\\label{2.n17a}
&&-f_bf_a\left(1+\varepsilon_a{f_a'h^3\over g^a_{\rm s}}\right)
\left(1+\varepsilon_b{f'_bh^3\over g^b_{\rm s}}
\right) \Biggr]\,F_{ba}\,\sigma_{ab}\,d\Omega_{ba}
{d^3p_b\over p_{b0}}.
\een
Above it was supposed that the external force that acts on the particles of electric charge $\q_a$ is of electromagnetic nature. In this case the Minkowski  force reads
\be
K_a^\alpha={\q_a\over c}F^{\alpha\beta}{p_{a\beta}\over
m_a},
\ee{8.128}
where $F^{\alpha\beta}$ is the electromagnetic field tensor.

Now we introduce  the moments of the distribution function, which are the partial particle four-flow $N_a^\alpha$ and the
partial energy-momentum tensor $T_a^{\alpha\beta}$. They are defined through:
\be
N_a^\alpha=c\int p_a^\alpha f_a {d^3 p_a\over p_{a0}},
\qquad T_a^{\alpha\beta}=c\int
p_a^\alpha p_a^\beta f_a
{d^3 p_a\over p_{a0}}.
\ee{8.131}
The corresponding quantities for the mixture read
\be
N^\alpha=\sum_{a=1}^r N^\alpha_a,
\qquad
T^{\alpha\beta}=\sum_{a=1}^r T_a^{\alpha\beta}.
\ee{8.132}

In the analysis of ionized gases it is also important  to introduce the electric charge
four-vector $J^\alpha$, which is defined in terms of the partial particle four-flows
$N^\alpha_a$ and of the partial electric charges $\q_a$ as
\be
J^\alpha=\sum_{a=1}^r \q_a N_a^\alpha.
\ee{8.133}

The balance equations for the particle four-flow and of the energy-momentum tensor of the mixture are obtained by multiplying  (\ref{2.n17a}) by $c$ and $cp_a^\a$, respectively, and by summing the resulting equations, yielding
\be
\partial_\alpha N^\alpha=0,\qquad \partial_\beta T^{\alpha\beta}={1\over c} F^{\alpha\beta} \sum_{a=1}^r \q_a
N_{a\beta}={1\over c} F^{\alpha\beta} J_\beta.
\ee{8.134}
Equation (\ref{8.134})$_1$ is the conservation law for the particle four-flow of the mixture. Equation (\ref{8.134})$_2$  when compared with the balance equation for
the energy-momentum tensor of the electromagnetic field $T_{\rm em}^{\alpha\beta}$
has an opposite sign on its right-hand side. However, if we denote the energy-momentum tensor of
(\ref{8.134})$_2$ by an index $\rm pt$ -- that refers to the particles -- we get
 the conservation law  (see Landau and Lifshitz~\cite{Landau8}):
\be
\partial_\alpha(T_{\rm pt}^{\alpha\beta}+
T_{\rm em}^{\alpha\beta})=0,
\ee{8.136}
which means that the sum of the energy-momentum tensors of
the particles and of the electromagnetic field satisfies a conservation equation.

\section{ Landau-Lifshitz decomposition}

The decomposition of the partial particle four-flow and of the partial
energy-momentum tensor proceeds by  introducing the four-velocity $U^\alpha$ and
the projector $\Delta^{\alpha\beta}$ defined by
\be
\Delta^{\alpha\beta}=\eta^{\alpha\beta}-{1\over c^2}U^\alpha U^\beta, \quad
\hbox{such that}\quad \Delta^{\alpha\beta}U_\beta=0.
\ee{a}

In the  Landau-Lifshitz  description \cite{LL} the
partial particle four-flow and the partial energy-momentum tensor may be decomposed according to
\ben\label{d1}
&&N_a^\alpha=n_a U^\alpha+J_a^\alpha-{n_aq^\alpha\over nh},
\\\nonumber
&&T_a^{\alpha\beta}=p_a^{\langle\alpha\beta\rangle}-(p_a+\varpi_a)
\Delta^{\alpha\beta}+{1\over c^2}
U^\alpha\left(q_a^\beta+h_a J_a^\beta-{n_ah_a\over nh}q^\beta\right)
\\\label{d2}
&&+{1\over c^2}
U^\beta\left(q_a^\alpha+h_a J_a^\alpha-{n_ah_a\over nh}q^\alpha\right)
+{e_an_a\over c^2}U^\alpha U^\beta.
\een
Above we have introduced the following  quantities for the constituent $a$ in the mixture:  particle number density $n_a$,
diffusion flux  $J^\alpha_a$,  pressure deviator $p_a^{\langle\alpha\beta\rangle}$, pressure
$p_a$, non-equilibrium pressure $\varpi_a$, heat flux $q_a^\alpha$,   energy per particle $e_\a$ and
enthalpy per particle $h_a=e_a+p_a/n_a$.
The corresponding quantities for the mixture are given by the sums
\ben\label{ff1}
n=\sum_{a=1}^r n_a,\qquad p^{\langle\alpha\beta\rangle}=\sum_{a=1}^rp_a^{\langle\alpha\beta\rangle},
\qquad
p=\sum_{a=1}^r p_a,\quad \varpi=\sum_{a=1}^r\varpi_a,\\\label{ff2}
 ne=\sum_{a=1}^r n_ae_a, \qquad q^\alpha=\sum_{a=1}^r(q_a^\alpha+h_aJ_a^\alpha),\qquad
n h=\sum_{a=1}^r n_a h_a.
\een

The sum of (\ref{d1}) and (\ref{d2}) over all constituents of the mixture  lead to the following decompositions of the particle four-flow and energy-momentum tensor of the mixture
\ben
&&N^\alpha=nU^\alpha-{q^\alpha\over h},\qquad
T^{\alpha\beta}=p^{\langle\alpha\beta\rangle}-(p+\varpi)\Delta^{\alpha\beta}+
{en\over c^2}U^\alpha U^\beta,
\een
thanks to the constraint that there exist  only $(r-1)$ partial diffusion fluxes that are linearly independent for a mixture
of $r$ constituents, namely,
\be
\sum_{a=1}^{r} J_a^\alpha=0.
\ee{8.140}

We may also define the electric current four-vector
$I^\alpha$  in terms of the partial diffusion
fluxes $J_a^\alpha$ and of the partial electric
charges $q_a$ as
\be
I^\alpha=\sum_{a=1}^r \q_a J_a^\alpha.
\ee{8.140a}

We refer to the works of de Groot and Suttorp~\cite{GRSu}
and of van Erkelens and van Leeuwen~\cite{vanvan} and
decompose the electromagnetic field tensor $F^{\alpha\beta}$ into one part
which is parallel to the four-velocity $U^\alpha$ and another
which is perpendicular to it, i.e.
\be
F^{\alpha\beta}={1\over c^2}\left(F^{\alpha\gamma}U_{\gamma} U^\beta-
F^{\beta\gamma}U_{\gamma} U^\alpha \right)+\Delta^\alpha_{\gamma}
F^{\gamma\delta}\Delta^\beta_{\delta}.
\ee{8.145}
Furthermore,   by introducing the tensors $E^\alpha$ and $B^{\alpha\beta}$ defined by
\be
E^\alpha={1\over c}F^{\alpha\beta}U_{\beta},\qquad
B^{\alpha\beta}=-\Delta^\alpha_{\gamma} F^{\gamma\delta}
\Delta^\beta_{\delta},
\ee{8.146}
we may write  the electromagnetic field
tensor  as
\be
F^{\alpha\beta}={1\over c}\left(E^\alpha U^\beta-
E^\beta U^\alpha \right)-B^{\alpha\beta}.
\ee{8.147}

If we consider  a local Lorentz rest frame where
$(U^\alpha)=(c,{\bf 0})$, equations
(\ref{8.146})  imply that
\be
(E^\alpha)=(0,{\bf E}),\qquad
B^{0\alpha}=B^{\alpha 0}=0,\qquad
B^{ij}=-c\epsilon^{ijk}B_k,
\ee{8.148}
and  we can identify $E^\alpha$ with the electric
field $\bf E$ and $B^{\alpha\beta}$
with  the magnetic flux induction  $\bf B$.

Due to the fact that  $F^{\alpha\beta}$ is an antisymmetric tensor
$F_{\alpha\beta}U^\alpha U^\beta=0$, it follows from (\ref{8.146}) and
(\ref{8.147}) the relationships
\be
E_\alpha U^\alpha=0,\qquad B_{\alpha\beta}U^\beta=0,\qquad
\hbox{and}\qquad B^{\alpha\beta}=-B^{\beta\alpha}.
\ee{8.149}

\section{Chapman-Enskog method}

Since we are interested  to derive the laws of
Fourier and Ohm for a binary mixture of electrons and protons and of
electrons and photons, we have to made some simplifications of our model, which are enumerated below:
\begin{enumerate}
\item the electric current four-vector (\ref{8.140a}) for a binary mixture of electrons $(a= e)$ and
protons $(a=p)$ may be written as
\be
I^\alpha=-2{\rm e}J^\alpha_e,
\ee{8.154}
since the relationship
between the diffusion fluxes reads
$J_e^\alpha=-J_p^\alpha$ and the electric charges are given
by $\q_e=-{\rm e}$, $\q_p={\rm e}$,  with ${\rm e}$ denoting
 the elementary charge.  Furthermore, we shall analyze
the so-called Lorentzian plasma~\cite{LifP} where
the collisions between the electrons may be neglected in
comparison with the collisions between the
electrons and protons. A Lorentzian plasma must
fulfill the condition that the
mass of one constituent is much larger than the mass of the other
constituent. Here we have that $m_p/m_e\approx 1836$,
where $m_e$ and $m_p$ denote the electron and proton masses, respectively.
Moreover, we shall assume a locally neutral system where
$\q_en_e+\q_pn_p=0$, which implies  that $n_e=n_p$;

\item the electric current four-vector
(\ref{8.140a}) for a binary mixture of electrons $(a=e)$ and
photons $(a=\gamma)$, reduces to
\be
I^\alpha=-{\rm e}J^\alpha_e,
\ee{8.155}
due to the fact that the electric charge of the
photons is zero $(\q_\gamma=0)$. Furthermore,
the collisions between electrons can also be neglected in
comparison to the collisions between electrons and photons,
which is the Compton scattering;

\item the partial heat fluxes of the protons and of the photons are negligible
in comparison with the partial heat flux of the electrons so that we can
write from (\ref{ff2})$_2$ that the heat flux of the mixture reduces to
\be
q^\alpha=q_e^\alpha+(h_e-h_b)J^\alpha_e,\qquad \hbox{with} \qquad b=p,\gamma.
\ee{8.144a}
\end{enumerate}

For simplicity we shall adopt the
Anderson and Witting model equation~\cite{AW} for the
electrons instead of using the relativistic Uehling-Uhlenbeck equation (\ref{2.n17a}). Hence, by taking into account the above considerations we write the space-time evolution of the
distribution function for the electrons as
\be
p_e^\alpha {\partial f_e\over \partial x^\alpha}
-{{\rm e}\over c}F^{\alpha\beta}p_{e\beta}{\partial f_e \over \partial p_e^\alpha}
=-{U^\alpha p_{e\alpha}\over c^2\tau_{eb}}(f_e-f_e^{(0)}),
\ee{8.156}
where $\tau_{eb}$ with $b=p$ or $b=\gamma$ is the mean free time between collisions of
electrons-protons or electrons-photons, respectively. In the above equation
$f_e^{(0)}$ is the equilibrium distribution function of the electrons which
reads
\be
f_e^{(0)}={2\over h^3}{1\over \exp\left({-{\mu_e\over kT}
+{U^\alpha p_{e\alpha}\over kT}}\right)+ 1},
\ee{8.76a}
by considering that the electrons obey the Fermi-Dirac statistics. Above,
 $T$ denotes the temperature of the mixture, $\mu_e$  the chemical potential of the electrons and the factor 2
refers to the degeneracy factor of the electrons.

Once we know the equilibrium distribution function of the electrons we may calculate the values of the fields at equilibrium: particle number density $n_e$, energy density $n_e e_e$ and pressure  $p_e$ defined by
\ben\label{nn1}
&&n_e=\frac{1}{c^2} U_\a N_e^\a=\frac{1}{c^2} U_\a\, c\int p_e^\a f_e^{(0)}\frac{d^3 p_e}{p_{e0}},\\
&&n_e e_e=\frac{1}{c^2} U_\a U_\b T_e^{\a\b}=\frac{1}{c^2} U_\a U_\b \,c\int p_e^\a p_e^\b f_e^{(0)}\frac{d^3 p_e}{p_{e0}},\\
&&p_e=-\frac{1}{3} \Delta_{\a\b} T_e^{\a\b}=-\frac{1}{3} \Delta_{\a\b} \,c\int p_e^\a p_e^\b f_e^{(0)}\frac{d^3 p_e}{p_{e0}}.
\een
The calculation proceeds as follows: we consider a local Lorentz rest system where $U^\a=(c,\bf0)$ so that the particle number density of the electrons (\ref{nn1}) reduces to
\ben\label{ne}
n_e=\int {2\over h^3}{1\over \exp\left({-{\mu_e\over kT}
+{c p_{e0}\over kT}}\right)+ 1} \vert \bp_e\vert^2 \sin\psi d\chi d\psi d \vert\bp_e\vert,
\een
where we have introduced the spherical coordinates $0\leq\psi\leq\pi$, $0\leq\chi\leq2\pi$ and $0\leq\vert\bp_e\vert<\infty$. Now we change the integration variable by introducing a new variable $\vartheta$ defined through
\ben
\vert\bp_e\vert = m_ec\sinh\vartheta,\qquad \hbox{so that}\qquad {c p_{e0}\over kT}=\z_e \cosh\vartheta,
\een
where $\z_e=m_ec^2/kT$ is the ratio between the electron rest  energy $m_ec^2$ and the thermal energy of the gas $kT$. When $\z_e\gg1$ the electron behaves as a non-relativistic gas, while when $\z_e\ll1$ it behaves as an ultra-relativistic gas. The change of variables and the integration of (\ref{ne}) in the angles $\chi$ and $\psi$ leads to
\ben
n_e=8\pi\left(\frac{m_ec}{h}\right)^3\int_0^\infty \frac{\sinh^2 \vartheta\cosh\vartheta d\vartheta}{\exp(-\mu_e^\star+\z_e\cosh\vartheta)+1}=\frac{8\pi}{h^3}\left(m_ec\right)^3\mathcal{J}_{21}(\z_e,\mu_e^\star).
\een
In the above equation we have introduced the electron chemical potential $\mu_e^\star=\mu_e/kT$ in units of $kT$ and the integral $\mathcal{J}_{nm}(\z_e,\mu_e^\star)$ defined by
\ben\label{int}
\mathcal{J}_{nm}(\z_e,\mu_e^\star)=\int_0^\infty \frac{\sinh^n \vartheta\cosh^m\vartheta d\vartheta}{\exp(-\mu_e^\star+\z_e\cosh\vartheta)+1}.
\een
Following the same methodology we get that
\ben
n_e e_e=\frac{8\pi}{h^3}m_e^4c^5\mathcal{J}_{22}(\z_e,\mu_e^\star),\qquad
p_e=\frac{8\pi}{h^3}m_e^4c^5\mathcal{J}_{40}(\z_e,\mu_e^\star).
\een

Now we shall  determine from  (\ref{8.156}) the
non-equilibrium distribution function for the electrons by adopting the Chapman-Enskog methodology. For that purpose we
search for a solution  of the form
\be
f_e=f_e^{(0)}+\phi_e,
\ee{d}
where the deviation from the equilibrium distribution function is
considered to be a small quantity, i.e., $\vert\phi_e\vert\ll1$. If we insert (\ref{d}) into the
Boltzmann equation (\ref{8.156}) we get
\be
p_e^\alpha {\partial f_e^{(0)}\over \partial x^\alpha}
-{{\rm e}\over c}F^{\alpha\beta}p_{e\beta}{\partial f_e^{(0)} \over \partial
p_e^\alpha}-{{\rm e}\over c}F^{\alpha\beta}p_{e\beta}{\partial \phi_e
\over \partial p_e^\alpha}=-{U^\alpha p_{e\alpha}\over c^2\tau_{eb}}\phi_e,
\ee{e}
where we have not taken into account the term ${\partial \phi_e/ \partial x^\alpha}$,
since it is not our aim  in deriving constitutive equations
which are functions of second-order derivatives (Burnett equations). The
above equation can be written as
\ben\nonumber
&&{-2\over h^3}{\exp\left({-{\mu_e\over kT}
+{U^\alpha p_{e\alpha}\over kT}}\right)\over \left[\exp\left({-{\mu_e\over kT}
+{U^\alpha p_{e\alpha}\over kT}}\right)+ 1\right]^2}\left\{{1\over c^2}
(p_e^\alpha U_\alpha)\left[D\left({\mu_e\over kT}\right)+{p_e^\beta
U_\beta\over kT^2}DT\right]\right.
\\\nonumber&&\left.
+{p_e^\beta U_\beta\over kT^2}p_{e\alpha}\left[\nabla^\alpha T
-{T\over c^2}DU^\alpha\right]-{{\rm e}\over kT}p_{e\alpha}\left[E^\alpha
 -{kT\over {\rm e}}
\nabla^\alpha \left({\mu_e\over kT}\right)
\right]\right.
\\\label{1}
&&\left.
-{p_{e\alpha} p_{e\beta}\over kT}\nabla^\alpha U^\beta
\right\}
={U^\gamma p_{e\gamma}\over c^2\tau_{eb}}\left[ 1+{m_ec\over
U^\delta p_{e\delta}}\left({\omega_e\tau_{eb}\over B}\right)B^{\alpha\beta}
p_{e\beta}{\partial\over \partial p_e^\alpha}\right]\phi_e,
\een
where we have not considered the term $E^\alpha{\partial\phi_e/\partial
p_e^\alpha},$ since it refers also to a second-order term. Furthermore, we have
introduced in the above equation the electron cyclotron frequency
$\omega_e={\rm e}B/m_e$ -- where $B$ is the modulus of the magnetic
flux induction -- and the differential operators $D\equiv U^\alpha\partial_\alpha$ and
$\nabla^\alpha\equiv\Delta^{\alpha\beta}\partial_\beta$.

In this work we are interested in the derivation of the laws of Fourier and Ohm, so that we can
restrict ourselves to the  thermodynamic forces that are four-vectors, namely
\be
\nabla^\alpha{\cal T}\equiv \left[\nabla^\alpha T-{T\over c^2}DU^\alpha\right]
\qquad\hbox{and}\qquad
{\cal E}^\alpha\equiv \left[E^\alpha-{kT\over {\rm e}}
\nabla^\alpha \left({\mu_e\over kT}\right)
\right],
\ee{6a}
the first being a combination of a temperature gradient and an
acceleration, while the second refers to a combination of an external electric
field and a gradient of the chemical potential of the electrons.
Hence, we obtain from (\ref{1}) that the deviation from the distribution
function may be written as
\be
\phi_e=A_\alpha\left\{{p_e^\beta U_\beta\over kT^2}\nabla^\alpha{\cal T}
-{{\rm e}\over kT}{\cal E}^\alpha\right\}.
\ee{2}
Up to terms in $(\omega_e\tau_{eb}/B)^2$  the four-vector $A^\alpha$  is given  by
\ben\nonumber
&&A^\alpha={-2\over h^3}{\exp\left({-{\mu_e\over kT}
+{U^\alpha p_{e\alpha}\over kT}}\right)\over \left[\exp\left({-{\mu_e\over kT}
+{U^\alpha p_{e\alpha}\over kT}}\right)+ 1\right]^2}{c^2\tau_{eb}\over
U^\gamma p_{e\gamma}}\left[\eta^{\alpha\beta}-{m_ec\over U^\delta
p_{e\delta}}\left({\omega_e\tau_{eb}\over B}\right)B^{\alpha\beta}\right.
\\\label{3}
&&\left.+\left({m_ec\over U^\delta p_{e\delta}}\right)^2\left({\omega_e\tau_{eb}
\over B}\right)^2B^{\alpha}_\gamma B^{\gamma\beta}\right]p_{e\beta}.
\een

Equation (\ref{2}) together with (\ref{3}) represent the deviation of the
distribution function of the electrons as a function of thermodynamic forces
that are four-vectors.  We shall use the distribution function (\ref{d}) in the following section
 in order to determine the laws of Ohm and Fourier.

\section{Ohm and Fourier laws}

The  determination of  the diffusion flux $J_e^\alpha$ and the heat flux
$q_e^\alpha$ of the electrons proceeds by noting that  (\ref{8.131}), (\ref{d1}) and
(\ref{d2}) lead to
\be
{h_b\over h}J_e^\alpha-{n_e\over nh}q_e^\alpha=\Delta^\alpha_\beta
N_e^\beta=\Delta^\alpha_\beta \int c
p_e^\beta f_e{d^3p_e\over p_{e0}},
\ee{4}
\be
{h_bh_e\over h}J_e^\alpha+{n_bh_b\over nh}q_e^\alpha=
\Delta^\alpha_\beta U_\gamma T_e^{\beta\gamma}=
\Delta^\alpha_\beta U_\gamma\int c
p_e^\beta p_e^\gamma f_e{d^3p_e\over p_{e0}}.
\ee{5}
The  insertion of the distribution function of the electrons (\ref{d}) together
with (\ref{2}) and (\ref{3}) into (\ref{4}) and (\ref{5}) and integration of the
resulting equations, implies a system of equations for $J_e^\alpha$ and
$q_e^\alpha$ which is used to determine the heat flux of the mixture
(\ref{8.144a}) and the electric current four-vector (\ref{8.154}) or
(\ref{8.155}). From this system of equations it follows Fourier and Ohm laws
\be
q^\alpha=\Lambda^{\alpha\beta}\nabla_\beta {\cal T}+
\Upsilon^{\alpha\beta}{\cal E}_\beta,
\qquad\qquad
I^\alpha=\sigma^{\alpha\beta}{\cal E}_\beta+\Omega^{\alpha\beta}
\nabla_\beta {\cal T},
\ee{6}
respectively. Above  $\Lambda^{\alpha\beta}$ is a tensor
associated with
the thermal conductivity, $\sigma^{\alpha\beta}$ is the
electrical conductivity tensor, while
the tensors $\Upsilon^{\alpha\beta}$ and $\Omega^{\alpha\beta}$ are
related with cross effects. We may represent the general expressions for the above
mentioned tensors as
\ben\nonumber
\{\Lambda^{\alpha\beta},\Upsilon^{\alpha\beta},\sigma^{\alpha\beta},
\Omega^{\alpha\beta}\}&=&\{a_1,b_1,c_1,d_1\}\eta^{\alpha\beta}
+\{a_2,b_2,c_2,d_2\}B^{\alpha\beta}\\
&+&\{a_3,b_3,c_3,d_3\}
B^{\alpha\gamma}B_\gamma^\beta,
\een
where the scalar coefficients $a_1$ through $d_3$ are given below:
\begin{enumerate}
\item Coefficients associated with $\Lambda^{\alpha\beta}$
\ben
a_1&=&{8\pi m_e^5 c^9\tau_{eb}h\over 3h^3h_b kT^2}
 \left(\mathcal{J}_{41}^\bullet-
{h_b\over  m_ec^2}\mathcal{J}_{40}^\bullet\right),
\\
a_2&=&{8\pi m_e^5c^9\tau_{eb}h\over 3h^3h_bkT^2}
\left(\mathcal{J}_{40}^\bullet-
{h_b\over  m_ec^2}\mathcal{J}_{4-1}^\bullet\right)\left({\omega_e\tau_{eb}\over cB}
\right),
\\
a_3&=&{8\pi m_e^5c^9\tau_{eb}h\over 3h^3h_bkT^2}
\left(\mathcal{J}_{4-1}^\bullet-
{h_b\over  m_ec^2}\mathcal{J}_{4-2}^\bullet\right)\left({\omega_e\tau_{eb}\over cB}
\right)^2.
\een

\item Coefficients associated with $\Upsilon^{\alpha\beta}$
\ben
b_1&=&-{8\pi m_e^4c^7\tau_{eb}h{\rm e}\over 3h^3h_bkT}
\left(\mathcal{J}_{40}^\bullet-
{h_b\over  m_ec^2}\mathcal{J}_{4-1}^\bullet\right),
\\
b_2&=&-{8\pi m_e^4c^7\tau_{eb}h{\rm e}\over 3h^3h_bkT}
\left(\mathcal{J}_{4-1}^\bullet-
{h_b\over  m_ec^2}\mathcal{J}_{4-2}^\bullet\right)\left({\omega_e\tau_{eb}\over cB}
\right),
\\
b_3&=&-{8\pi m_e^4c^7\tau_{eb}h{\rm e}\over 3h^3h_bkT}
\left(\mathcal{J}_{4-2}^\bullet-
{h_b\over  m_ec^2}\mathcal{J}_{4-3}^\bullet\right)
\left({\omega_e\tau_{eb}\over cB}\right)^2.
\een

\item Coefficients associated with $\sigma^{\alpha\beta}$
\ben
c_1&=&{8\pi m_e^4c^7\tau_{eb}n_e(Z+1){\rm e}^2\over
3h^3nh_bkT} \left(\mathcal{J}_{40}^\bullet+
{n_bh_b\over n_e m_ec^2}\mathcal{J}_{4-1}^\bullet\right),
\\
c_2&=&{8\pi m_e^4c^7\tau_{eb}n_e(Z+1){\rm e}^2\over
3h^3nh_bkT}  \left(\mathcal{J}_{4-1}^\bullet+
{n_bh_b\over n_e m_ec^2}\mathcal{J}_{4-2}^\bullet\right)\left({\omega_e\tau_{eb}
\over cB}\right),
\\
c_3&=&{8\pi m_e^4c^7\tau_{eb}n_e(Z+1){\rm e}^2\over
3h^3nh_bkT}\left(\mathcal{J}_{4-2}^\bullet+
{n_bh_b\over n_e m_ec^2}\mathcal{J}_{4-3}^\bullet\right)\left({\omega_e\tau_{eb}
\over cB}\right)^2.
\een

\item Coefficients associated with $\Omega^{\alpha\beta}$
\ben
d_1&=&-{8\pi m_e^5c^9\tau_{eb}n_e(Z+1){\rm e}\over
3h^3nh_bkT^2}  \left(\mathcal{J}_{41}^\bullet+
{n_bh_b\over n_e m_ec^2}\mathcal{J}_{40}^\bullet\right),
\\
d_2&=&-{8\pi m_e^5c^9\tau_{eb}n_e(Z+1){\rm e}\over
3h^3nh_bkT^2}
\left(\mathcal{J}_{40}^\bullet+
{n_bh_b\over n_e m_ec^2}\mathcal{J}_{4-1}^\bullet\right)  \left({\omega_e\tau_{eb}
\over cB}\right),
\\
d_3&=&-{8\pi m_e^5c^9\tau_{eb}n_e(Z+1){\rm e}\over
3h^3nh_bkT^2}  \left(\mathcal{J}_{4-1}^\bullet+
{n_bh_b\over n_e m_ec^2}\mathcal{J}_{4-2}^\bullet\right)\left({\omega_e\tau_{eb}
\over cB}\right)^2.
\een
\end{enumerate}
In the above equations  $\mathcal{J}_{nm}^\bullet$ represents the
partial derivative of (\ref{int})
with respect to the chemical potential of the electrons
$\mu_e^\star=\mu_e/(kT)$
in units of $kT$. Furthermore, we have introduced the abbreviation
$\zeta_e=m_ec^2/(kT)$
which refers to the ratio between the rest energy of the electrons
$m_ec^2$ and
the thermal energy of the mixture $kT$. We note
 that in all above equations one has to consider $Z=1$ for binary
mixtures of electrons and protons and $Z=0$ for binary mixtures of electrons
and photons.

The thermal conductivity tensor $\lambda^{\alpha\beta}$
is obtained by eliminating ${\cal E}^\alpha$ from (\ref{6})$_1$ through the use of (\ref{6})$_2$  by assuming that there is no
electric current. Hence, we get a relationship between
${\cal E}^\alpha$ and $\nabla^\alpha {\cal T}$ from (\ref{6})$_2$
which may be used to write Fourier law as
\be
q^\alpha=\lambda^{\alpha\beta}\nabla_\beta {\cal T},\quad\hbox{where}\quad
\lambda^{\alpha\beta}=e_1\eta^{\alpha\beta}+e_2B^{\alpha\beta}+e_3
B^{\alpha\gamma}B_\gamma^\beta.
\ee{7b}
 Up to terms in $[\omega_e\tau_{eb}/(cB)]^2$ the scalar coefficients $e_1$ through $e_3$
 read
\ben
e_1={a_1c_1-b_1d_1\over c_1},\qquad
e_2={a_2c_1^2-b_1(c_1d_2-c_2d_1)-b_2c_1d_1\over c_1^2},
\\
e_3={a_3c_1^3-b_1[d_1(c_2^2-c_1c_3)-c_1c_2d_2]-c_1^2(b_1d_3+b_3d_1)
-c_1b_2(c_1d_2-c_2d_1)\over c_1^3}.
\een

In order to get a better physical interpretation of the components of the
tensors, it is usual in the theory of ionized gases
to decompose the thermodynamic forces $\nabla_\alpha{\cal T}$
and ${\cal E}_\alpha$ into parts
parallel, perpendicular and
transverse to the magnetic flux induction. To achieve this goal we follow van Erkelens
and van Leeuwen~\cite{vanvan} and introduce the dual $\tilde B^{\alpha\beta}$
of the magnetic flux induction tensor
$B^{\alpha\beta}$ defined by
\be
\tilde B^{\alpha\beta}={1\over 2}\epsilon^{\alpha\beta\gamma\delta}B_{\gamma
\delta}.
\ee{8.171}
One may easy  verify from (\ref{8.171}) and (\ref{8.148}) that in a local
Lorentz rest frame the only non-zero
components of $\tilde B^{\alpha\beta}$ are
$\tilde B^{0i}=cB^i$ since $\tilde B^{ij}=0$ and $\tilde B^{00}=0$.

The desired decomposition of the thermodynamic forces into
parallel $\nabla^\alpha_\parallel{\cal T}$,
${\cal E}^\alpha_\parallel$; perpendicular
$\nabla^\alpha_\perp{\cal T}$, ${\cal E}^\alpha_\perp$ and
transverse $\nabla^\alpha_t{\cal T}$,
${\cal E}^\alpha_t$ parts read
\ben\label{8.172}
{\cal F}^\alpha_\parallel&=&{1\over \left({1\over 2}B^{\gamma\delta}
B_{\gamma\delta}\right)}\tilde B^{\alpha\beta}\tilde B_{\beta\gamma}
{\cal F}^{\gamma},
\qquad
{\cal F}^\alpha_\perp={-1\over \left({1\over 2}B^{\gamma\delta}
B_{\gamma\delta}\right)}B^{\alpha\beta}B_{\beta\gamma}
{\cal F}^{\gamma},
\\\label{8.174}
{\cal F}^\alpha_t&=&{1\over \left({1\over 2}B^{\gamma\delta}
B_{\gamma\delta}\right)^{1\over 2}}B^{\alpha\beta}
{\cal F}_{\beta},
\een
where ${\cal F}^\alpha$ is an abbreviation for  ${\cal E}^\alpha$ or $\nabla^\alpha{\cal T}$.
In a local Lorentz rest frame (\ref{8.172})
and (\ref{8.174}) reduce to
\ben
{\cal F}^0_\parallel={\cal F}^0_\perp={\cal F}^0_t=0,
\qquad
\mbox {\boldmath$\cal F$}_\parallel={1\over B^2}({\bf B}\cdot
\mbox {\boldmath$\cal F$}){\bf B},
\\
\mbox {\boldmath$\cal F$}_\perp={1\over B^2}[({\bf B}\cdot
\mbox {\boldmath$\cal F$}){\bf B}-({\bf B}\cdot{\bf B})
\mbox {\boldmath$\cal F$}],\qquad
\mbox {\boldmath$\cal F$}_t={1\over B}(
\mbox {\boldmath$\cal F$}\times{\bf B}),
\een
thanks to the relationship $\sqrt{B^{\gamma\delta}B_{\gamma\delta}/2}=c\sqrt{{\bf B}\cdot
{\bf B}}=cB$. From the above equations it is easy to
verify that $\mbox {\boldmath$\cal F$}_\parallel$ is parallel to the
magnetic flux induction $\bf B$,
$\mbox {\boldmath$\cal F$}_\perp$
perpendicular to it while $\mbox {\boldmath$\cal F$}_t$ is perpendicular to
both $\mbox {\boldmath$\cal F$}_\parallel$ and $\mbox {\boldmath
$\cal F$}_\perp$.

Now by using the following relationship
\be
(cB)^2\eta^{\alpha\beta}=\tilde
B^{\alpha\gamma}\tilde B_\gamma^{\;\beta}-B^{\alpha\gamma}B_\gamma^{\;\beta},
\ee{8.177}
the Fourier and Ohm laws can be rewritten  in terms of
${\cal F}^\alpha_\parallel$, ${\cal F}^\alpha_\perp$ and  ${\cal F}^\alpha_t$.
In fact, if we substitute (\ref{8.177}) into the Ohm's law (\ref{6})$_2$
and Fourier's law (\ref{7b})$_1$ and
make use of the definitions (\ref{8.172}) and (\ref{8.174}), it follows that
 the electric current four-vector and the heat flux can be written, without the cross-effects terms,  as
\be
I^\alpha=\sigma_\parallel {\cal E}^\alpha_\parallel+\sigma_\perp
{\cal E}^\alpha_\perp+\sigma_t {\cal E}^\alpha_t,\quad
q^\alpha=\lambda_\parallel
\nabla^\alpha_\parallel{\cal T}+\lambda_\perp
\nabla^\alpha_\perp{\cal T}+\lambda_t\nabla^\alpha_t{\cal T},
\ee{8.178a}
respectively.
In the above equations the scalars  are called the
parallel, perpendicular and transverse components of the  tensors, and their
expressions are given by
\ben\label{8.178b}
\left\{
            \begin{array}{ll}
              \sigma_\parallel=c_1,\qquad \sigma_\perp=c_1-c_3(cB)^2,\qquad
\sigma_t=c_2(cB),\\
             \lambda_\parallel=e_1,\qquad
\lambda_\perp=e_1-e_3(cB)^2,\qquad
\lambda_t=e_2(cB).
\end{array}
            \right.
\een
From the above formulas we shall obtain the parallel, perpendicular and
transverse components of the electrical and thermal conductivities for
binary mixtures of electrons and protons and of electrons and photons.

\section{Electrical and thermal conductivities}

\subsection{Non-degenerate electrons}

 Here we shall analyze two important cases, namely:   a non-relativistic mixture of protons and
non-degenerate electrons
and  an ultra-relativistic mixture of photons and non-degenerate electrons.
We note that the chemical potential of the electrons in the non-degenerate case
must fulfill the condition that  $e^{-\mu_e^\star}\gg1$.
\begin{enumerate}

\item A non-relativistic mixture of electrons and protons is identified by
two conditions $m_p/m_e\gg1$ and $\zeta_e=m_ec^2/(kT)\gg1$. In this case the
transport coefficients  read
\ben\label{n1}
\sigma_\parallel={e^2\tau_{ep}n_e\over m_e }\left(1-{5\over 2\zeta_e}
\right),\qquad \sigma_t={e^2\tau_{ep}n_e(\omega_e\tau_{ep})
\over m_e }\left(1-{5\over \zeta_e}\right),
\\\label{n2}
\sigma_\perp={e^2\tau_{ep}n_e\over m_e }\left[\left(1-{5\over 2\zeta_e}
\right)-(\omega_e\tau_{ep})^2\left(1-{15\over 2\zeta_e}\right)\right],
\\\label{n3}
\lambda_\parallel={5k^2T\tau_{ep}n_en\over2 m_en_p }\left(1-{3\over \zeta_e}
\right),\qquad \lambda_t={5k^2T\tau_{ep}n_en(\omega_e\tau_{ep})
\over 2m_en_p}\left(1-{15\over 2\zeta_e}\right),
\\\label{n4}
\lambda_\perp={5k^2T\tau_{ep}n_en\over 2m_en_p }\left[\left(1-{3\over \zeta_e}
\right)-(\omega_e\tau_{ep})^2\left(1-{12\over \zeta_e}\right)\right].
\een
 The first relativistic corrections to the
transport coefficients are related to the term $1/\zeta_e$ and if fix our attention to the leading terms without the relativistic
corrections, the electrical conductivities can be written from (\ref{n1}) and  (\ref{n2})
as:
\be
\sigma_\parallel={e^2\tau_{ep}n_e\over m_e },\qquad
\sigma_t=\sigma_\parallel(\omega_e\tau_{ep}),
\qquad
\sigma_\perp\approx{\sigma_\parallel\over 1+(\omega_e\tau_{ep})^2},
\ee{n2a}
since we have considered $\omega_e\tau_{ep}\ll1$. The  expressions for the electrical conductivities (\ref{n2a}) are well-known  in the theory
of non-degenerate and non-relativistic ionized gases (see, for example,
Cap~\cite{Cap}) and show their dependence on the
magnetic flux induction $B$ through the electron cyclotron frequency $\omega_e$.
Furthermore, the thermal conductivities (\ref{n3}) and  (\ref{n4}) without the relativistic corrections
 become
\be
\lambda_\parallel={5k^2T\tau_{ep}n_en\over2 m_en_p },\qquad
\lambda_t=\lambda_\parallel(\omega_e\tau_{ep}),
\qquad
\lambda_\perp\approx{\lambda_\parallel\over 1+(\omega_e\tau_{ep})^2}.
\ee{n2b}
Note that the expression for the parallel thermal conductivity is well-known in the theory of
non-relativistic gases which follow from a BGK model equation.

\item An ultra-relativistic mixture of photons and non-degenerate electrons
is characterized by the condition  $\zeta_e=m_ec^2/(kT)\ll1$. Here  the
transport coefficients reduce to
\be
\sigma_\parallel=\sigma_\perp={e^2c^2\tau_{e\gamma}n_e(3n_e+4n_\gamma)\over
12nkT},
\quad \sigma_t={e^2c^2\tau_{e\gamma}n_e(n_e+2n_\gamma)(\omega_e\tau_{e\gamma})
\zeta_e\over 12nkT},
\ee{u1}
\be
\lambda_\parallel=\lambda_\perp={4kc^2\tau_{e\gamma}n_en\over 3n_e+4n_\gamma},
\qquad \lambda_t={8kc^2\tau_{e\gamma}n_en^2(\omega_e\tau_{e\gamma})\zeta_e
\over (3n_e+4n_\gamma)^2}.
\ee{u2}
We infer  from these equations that the parallel and perpendicular electrical and thermal conductivities
coincide, while the transverse electrical and thermal conductivities are small quantities since they
are proportional to $\zeta_e$.
\end{enumerate}

\subsection{Completely degenerate electrons}

All thermal conductivities vanish in the limit of completely degenerate electrons, since this behavior is connected with the well-known result
from statistical mechanics that the heat capacity  of a completely
degenerate gas vanishes. For the electrical conductivities  there exist  three important cases to be analyzed  which
are: a  non-relativistic mixture of protons and completely degenerate
electrons; an ultra-relativistic mixture of photons and completely
degenerate  electrons and a mixture of non-relativistic protons and
ultra-relativistic completely degenerate  electrons. We proceed to analyze
the electrical conductivities for these cases.

\begin{enumerate}

\item A  non-relativistic mixture of protons and completely
degenerate electrons is identified by $\zeta_e\gg1$ and $p_F\ll m_ec$, where
$p_F$ denotes the Fermi momentum of the electrons. Here we have
\ben
\sigma_\parallel={8\pi e^2\tau_{ep}p_F^3\over 3m_e h^3}\left(1-{p_F^2\over
2m_e^2c^2}
\right),\quad \sigma_t={8\pi e^2\tau_{ep}p_F^3(\omega_e\tau_{ep})\over
3m_e h^3}\left(1-{p_F^2\over m_e^2c^2}\right),
\\
\sigma_\perp={8\pi e^2\tau_{ep}p_F^3\over 3m_e h^3}\left[\left(1-{p_F^2
\over 2m_e^2c^2}
\right)-(\omega_e\tau_{ep})^2\left(1-{3p_F^2\over 2m_e^2c^2}
\right)\right].
\een
Let us fix our attention
to the leading terms of the electrical conductivities
\be
\sigma_\parallel={8\pi e^2\tau_{ep}p_F^3\over 3m_e h^3},\qquad
\sigma_t=\sigma_\parallel(\omega_e\tau_{ep}),
\qquad
\sigma_\perp\approx{\sigma_\parallel\over 1+(\omega_e\tau_{ep})^2},
\ee{n6a}
since the term $p_F/(m_ec^2)$ is a small quantity
and the condition $\omega_e\tau_{ep}\ll1$ holds. These equations show the
dependence of the electrical conductivities on the magnetic flux induction $B$ through
the electron cyclotron frequency $\omega_e$.

\item An ultra-relativistic mixture of photons and completely degenerate  electrons
is characterized by the conditions $\zeta_e\ll1$ and $p_F\gg m_ec$, and the electrical conductivities for this case read
\ben
\sigma_\parallel=\sigma_\perp={8\pi e^2\tau_{e\gamma}c^2n_ep_F^3\over
12nkTh^3 }\left(1+{4kTn_\gamma\over n_ecp_F}\right),
\\
\sigma_t={8\pi e^2\tau_{e\gamma}n_e\zeta_ec(\omega_e\tau_{e\gamma})p_F^2
\over12n h^3 }\left(1+{4kTn_\gamma\over n_ecp_F}\right).
\een
We infer from the above equations that the parallel and perpendicular
electrical conductivities are equal to each other, while the transverse electrical
conductivity is a small quantity since it is proportional to $\zeta_e$.

\item A mixture of non-relativistic protons and ultra-relativistic
completely degenerate  electrons is also an important case since
it could describe a white dwarf star. Here  the conditions
$m_p/m_e\gg1$ and $p_F\gg m_ec$ hold and the electrical conductivities become
\be
\sigma_\parallel=\sigma_\perp={8\pi e^2\tau_{ep} cp_F^2\over 3h^3 }
\left(1-{m_e^2c^2\over 2p_F^2}\right),\quad
\sigma_t={8\pi e^2\tau_{ep}\zeta_ekT(\omega_e\tau_{ep})p_F
\over3 h^3 },
\ee{u3}
showing that the parallel and perpendicular conductivities coincide and that
the transverse conductivity is a small quantity, since it is proportional to $\zeta_e\ll 1$.
\end{enumerate}

\section*{Appendix: Integrals $\mathcal{J}_{nm}^\bullet$}
\begin{enumerate}
\item
{\bf  Non-degenerate case}

In this case $e^{-\mu_e^\star}\gg1$ so that the integrals $\mathcal{J}_{nm}$ reduce to
\be
\mathcal{J}_{nm}(\zeta_e,\mu_e^\star)=\int_0^\infty e^{-\zeta_e\cosh\vartheta+
\mu_e^\star}\sinh^n\vartheta\cosh^m\vartheta d\vartheta,
\ee{AL2}
and the integrals $\mathcal{J}_{nm}^\bullet(\zeta_e,\mu_e^\star)\equiv J_{nm}^\bullet$
can be expressed in terms of the modified Bessel
functions of the second kind $K_n(\zeta_e)\equiv K_n$ and of their  integrals
Ki$_n(\zeta_e)\equiv$Ki$_n$ (see Abramowitz and Stegun~\cite{AS} pages
376 and 483) as follows:
\ben
\mathcal{J}_{41}^\bullet={e^{\mu_e^\star}\over 2\zeta_e}(K_4-K_2),\quad
\mathcal{J}_{40}^\bullet={3e^{\mu_e^\star}\over 4\zeta_e}(K_3-K_1),\quad
\mathcal{J}_{4-1}^\bullet={e^{\mu_e^\star}\over \zeta_e}(K_2-{\rm Ki}_2),
\\
\mathcal{J}_{4-2}^\bullet={e^{\mu_e^\star}\over \zeta_e}(K_1+{\rm Ki}_1-{\rm Ki}_3),
\qquad
\mathcal{J}_{4-3}^\bullet={3e^{\mu_e^\star}\over \zeta_e}({\rm Ki}_2-{\rm Ki}_4).
\een
Moreover, the chemical potential of the electrons is given by
\be
e^{\mu_e^\star}={n_eh^3\over 8\pi m_e^2ckTK_2}.
\ee{AL5}

\item
{\bf  Completely degenerate case}

The integrals $\mathcal{J}_{nm}$ for this case reduce to
\be
\mathcal{J}_{nm}=\int_0^{\vartheta_F}
\sinh^n\vartheta\cosh^m\vartheta d\vartheta, \quad\hbox{with}\quad
\vartheta_F={\rm arcosh}\sqrt{1+\left({p_F\over m_ec}\right)^2},
\ee{AL6}
where $p_F$ is the Fermi momentum of the electrons. The integrals
$\mathcal{J}_{nm}^\bullet$ read
\ben
\mathcal{J}_{41}^\bullet={1\over \zeta_e}\left({p_F\over m_ec}\right)^3\sqrt{1+
\left({p_F\over m_ec}\right)^2},\qquad \mathcal{J}_{4-1}^\bullet\!=\!{\left({p_F\over m_ec}\right)^3\over \zeta_e\sqrt{1+
\left({p_F\over m_ec}\right)^2}},
\\
\mathcal{J}_{40}^\bullet={1\over \zeta_e}\left({p_F\over m_ec}\right)^3,\;
\mathcal{J}_{4-2}^\bullet\!=\!{\left({p_F\over m_ec}\right)^3\over \zeta_e\left[1+
\left({p_F\over m_ec}\right)^2\right]},\;
\mathcal{J}_{4-3}^\bullet\!=\!{\left({p_F\over m_ec}\right)^3\over \zeta_e\left[{1+
\left({p_F\over m_ec}\right)^2}\right]^{3\over 2}}.
\een
\end{enumerate}
\begin{theacknowledgments}
 The paper was partially supported by Brazilian Research Council (CNPq).
 \end{theacknowledgments}

  \bibliographystyle{aipproc}


\end{document}